\def\beq{\begin{equation}}
\def\eeq{\end{equation}}
\documentclass{article}
\usepackage{graphics}
\begin{document}

\title{Pointlike constituent quarks and scattering equivalences}

\author{Evan Sengbusch,  W. N. Polyzou\footnote{This work supported 
in part by the U.S. Department of Energy, under contract DE-FG02-86ER40286}\\
Department of Physics and Astronomy \\
The University of Iowa, Iowa City, IA 52242
}
\maketitle
%

\begin{abstract}

In this paper scattering equivalences are used to simplify current
operators in constituent quark models.  The simplicity of the method
is illustrated by applying it to a relativistic constituent quark
model that fits the meson mass spectrum\cite{ckp}.  This model
requires a non-trivial constituent quark current operator to fit the
pion form factor data.  A model with a different confining
interaction, that has the identical spectrum and can reproduce the
measured pion form factor using only point-like constituent quark
impulse currents is constructed.  Both the original and transformed
models are relativistic direct-interaction models with a light-front
kinematic subgroup \cite{bkwp}\cite{fc}.

\end{abstract}


\section{Introduction}

The prominent role played by the choice of representation of the
dynamics is a striking feature of relativistic direct-interaction
quantum mechanics.  Models with different kinematic subgroups \cite{bkwp}
(light-front, Euclidean, or Lorentz) that fit the experimental meson
and baryon spectra require different representations of quark
current\cite{salme}\cite{allen}\cite{fcdr}\cite{bertrand1}
\cite{krutov} operators to fit the
measured electromagnetic properties of the pion and nucleon.

The representation dependence of operators, like the current, is not
surprising.  In local quantum field theory all fields in the same Borchers
class \cite{sw} have the same scattering matrix.  In effective
field theory, field redefinitions \cite{dick} lead to equivalent
representations of dynamics on limited energy scales, and in quantum
theories of a finite number of degrees of freedom the group of
scattering equivalences \cite{fcwp1}\cite{wp} leads to different 
representations of the dynamics with the same spectrum and 
scattering matrix elements.

If the inverse scattering problem has a solution in relativistic or
non-relativistic quantum mechanics then it cannot be unique.  A large
class of equivalent interactions that lead to the same $S$-matrix can
be constructed using scattering equivalences \cite{wp}.  Different scattering
equivalent models are equivalent representations of the dynamics.

While there is no fundamental preference for a given representation
among the class of equivalent representations, there are computational
advantages to choosing a representation where one aspect of the dynamics
is simple.

We illustrate how the choice of representation may be used to simplify
the structure of the quark current operator in a relativistic
constituent quark model of mesons.  The general construction is
discussed in section two.  The construction is applied to the model of
\cite{ckp} in section three.

\section{Construction} Consider a relativistic constituent quark model 
which fits the meson mass spectrum.  Assume that this model is a
relativistic direct interaction model with a given kinematic subgroup.
Also assume that there is a corresponding constituent quark current
operator $J^{\mu} (x)$ that can be used to compute the electromagnetic
properties of the mesons.  In general the current operator includes
quark form factors and dynamical two-body operators.

The current matrix elements are invariant under simultaneous unitary 
transformations of the meson eigenstates and current operator
\beq
\vert p, \phi \rangle \to
\vert p, \phi' \rangle := A \vert p, \phi \rangle
\eeq
\beq
\vert p', \psi \rangle \to
\vert p', \psi' \rangle := A \vert p', \psi \rangle
\label{eq:AAa}
\eeq
\beq
J^{\mu} (x) \to J^{\prime \mu} (x) := A J^{ \mu} (x) A^{\dagger}
\label{eq:ABa}
\eeq
\[
\Downarrow
\]
\beq
\langle p', \psi \vert J^{\mu} (x) \vert p, \phi \rangle =
\langle p', \psi' \vert J^{\prime \mu} (x) \vert p, \phi' \rangle 
\label{eq:ACa}
\eeq
where $A$ is any unitary transformation.  The freedom to choose $A$
can be used to simplify the structure of the current operator or the
meson wave functions.

This work utilizes a restricted class of unitary transformations
that differ from the identity by a finite rank operator on the
internal Hilbert space.  For a suitable choice of the finite rank
operator, the original and transformed mass operators differ by a
short range interaction. The constituent quark masses and kinetic
energy remain unchanged and the original and transformed current
operators will differ by a short range operator.  If the finite rank
operator commutes with the kinematic subgroup, then the original and
transformed interactions will be kinematically invariant.

Unitary operators that are finite rank perturbations of the identity
are elements of class of scattering equivalences \cite{wp}.
Scattering equivalences are unitary transformations that preserve the
scattering matrix without changing the representation of free
particles.  While general unitary transformations could be used in
models with confining interactions, the scattering equivalences have
the desirable feature that they lead to equivalent models that differ
from the original model by a short-ranged addition to the original
confining interaction.

These unitary transformations have the structure
\beq
A= I + \Delta 
\label{eq:AA}
\eeq
where $\Delta$ is the finite rank part of $A$.  The
constituent quark-antiquark mass operator and current operator
transform as:
\beq
M = M_0 +V_c \to 
M' = AMA^{\dagger} = M_0 + V_c' 
\label{eq:AB}
\eeq
where 
\beq
V_c' = V_c + \Delta M + M \Delta^{\dagger} + \Delta M \Delta^{\dagger}
\label{eq:AC}
\eeq
and
\[
J^{\mu}(x)   \to 
J^{\prime \mu}(x)   = A J^{\mu}(x) A^{\dagger} =
\]
\beq
J^{\mu} (x) +
\Delta J^{\mu} (x) + J^{\mu} (x) \Delta^{\dagger} + 
\Delta J^{\mu} (x) \Delta^{\dagger} .
\label{eq:AD}
\eeq
Since $M$ is unbounded the range of $\Delta^{\dagger}$ is
required to be in the domain of $M$.  With this choice the differences
$M'-M$ and $J^{\prime \mu} (x) - J^{\mu}(x)$ are short ranged. 

The mass operator $M$ 
has the structure:
\beq
M= M_0 + V_c
\label{eq:AL}
\eeq
where 
\beq
M_0 = \sqrt{m_q^2 + k^2} + \sqrt{m_{\bar{q}}^2 + k^2}
\label{eq:AM}
\eeq
and
\beq
V_c = V_c(\lambda_1, \cdots , \lambda_n)
\label{eq:AN}
\eeq 
is a phenomenological confining interaction with parameters
$\lambda_1, \cdots, \lambda_n$ chosen to fit the mass spectrum of 
observed mesons.

The eigenvectors and eigenvalues of $M$ are denoted by
\beq
M\vert m_n \rangle = m_n \vert m_n \rangle 
\label{eq:AO}
\eeq
where the $n=0$ state corresponds to the $\pi$ meson and
$m_0= m_{\pi}$. 

To generate the finite rank operator $\Delta$ let $\vert \bar{m_\pi}
\rangle$ be the lowest mass eigenstate of a mass operator $\bar{M}$
obtained from $M$ changing parameters $\{\lambda_1, \cdots,
\lambda_n\}$ in the confining interaction: 
\beq
\bar{M} = M_0+ \bar{V}_c  \qquad
\bar{M} \vert \bar{m}_\pi \rangle =
\bar{m}_\pi \vert \bar{m}_\pi \rangle  
\label{eq:AP}
\eeq
where 
\beq
\bar{V}_c = V_c(\bar{\lambda}_1, \cdots , \bar{\lambda}_n) .
\label{eq:AQ}
\eeq
Assume that it is possible to find a set of parameters 
$\{\bar{\lambda}_1, \cdots , \bar{\lambda}_n\}$ with the
property that experimental pion form factor can be expressed in terms
of a matrix element of a point-like impulse current, $J_p^{\mu} (0)$,
in the states $\vert p, \bar{m}_{\pi}, \rangle$, $\vert p', \bar{m}_{\pi}
\rangle$. 

The wave function $\langle k \vert \bar{m}_{\pi} \rangle$ is a
candidate for the transformed pion wave function.
The assumption means that the desired wave function can 
be found in the class of ground state solutions of mass operators with
different choices of the potential parameters.  This restricts the
possible candidates for the new wave function to a set of functions that
depend on a small number of parameters.  A larger class of wave
functions can be considered if there are no functions in the
above class that are able to reproduce the measured form factors.
The spectrum of $\bar{M}$ contains no physics, however,
without loss of generality one of the parameters,
$\bar{\lambda}_k$, can be taken to be an additive constant which 
can be adjusted so the eigenvalue $\bar{m}_{\pi} = m_{\pi}$.  This 
choice simplifies the notation, but has no effect on the form factor.

A general feature of relativistic quantum mechanical models is that
an impulse current cannot be compatible with current conservation
and current covariance.  The problem is that the left side of 
the relations
\beq
U(\Lambda ,a) J^{\mu} (x) U^{\dagger} (\Lambda,a) =
J^{\nu}(\Lambda x + a) \Lambda_{\nu} {}^{\mu}   
\eeq
and
\beq
[H,J^0 (x) ]_-= \sum_i [P^i , J^i (x) ]_- 
\eeq
involves the dynamics.  These dynamical constraints, when combined with 
space reflection and time reversal symmetries,  imply that the 
current matrix elements can be expressed in terms of a smaller set of
independent current matrix elements.  The number of independent
current matrix elements is precisely equal to the number of invariant
form factors.  For each kinematic subgroup there are natural choices
of independent current matrix elements.  It is possible to
consistently evaluate the independent current matrix elements, and all 
matrix elements related to them by kinematic transformations, using only
impulse currents.  In this paper, the impulse approximation means 
only that the independent matrix elements are calculated in the 
impulse approximation.   

The problem is to find a unitary operator $A=I+\Delta$ that maps
$\vert m_\pi \rangle$ to $\vert \bar{m}_\pi \rangle$.  Since $\vert
\bar{m}_\pi \rangle$ is not orthogonal to any of the states $\vert
m_n\rangle$, except where required by selection rules, the
transformation $A$ will effect all of the states.

A minimal solution uses a rank-one $\Delta$.  To construct the desired
transformation define two orthogonal bases on the two-dimensional
subspace spanned by the vectors $\vert m_\pi \rangle$ and $\vert
\bar{m}_\pi \rangle$.  If these vectors are identical then $A=I$, and there
is nothing to do.  Otherwise orthogonal bases on this 
two-dimensional subspace are
\beq
\vert m_\pi \rangle,
\qquad
\vert  m_\perp \rangle := {\vert \bar{m}_\pi \rangle
-  \vert m_\pi \rangle \cos (\theta)  \over \sin (\theta)} 
\label{eq:AS}
\eeq
and 
\beq
\vert \bar{m}_\pi \rangle, 
\qquad
\vert \bar{m}_\perp \rangle := {\vert {m}_\pi \rangle
-  \vert \bar{m}_\pi \rangle \cos (\theta)  \over \sin (\theta)} 
\label{eq:AT}
\eeq
where
\beq
\cos (\theta)= \cos (\theta)^* :=\langle m_\pi \vert \bar{m}_\pi \rangle
\label{eq:AU}
\eeq
\beq
\sin (\theta ) > 0 .
\label{eq:AV}
\eeq
The overlap $\langle m_\pi \vert \bar{m}_\pi \rangle = \cos (\theta)$ 
can be chosen to be real as a consequence of the time reversal
invariance of $V_c$ and $\bar{V}_c$.  The $\sin (\theta)$ is non zero,
otherwise $\vert m_{\pi} \rangle = \vert \bar{m}_{\pi} \rangle$.

By construction the states 
$\{ \vert m_\pi \rangle ,\vert \bar{m}_\pi \rangle\}$,
$\{ \vert m_\pi \rangle ,\vert m_\perp \rangle\}$, and 
$\{ \vert \bar{m}_\pi \rangle ,\vert \bar{m}_\perp \rangle\}$
all span the same two-dimensional subspace.  

The operator $A$ is chosen to satisfy 
\beq
A \vert m_\pi \rangle = \vert \bar{m}_\pi \rangle 
\label{eq:AW}
\eeq
\beq
A\vert {m}_\perp \rangle = \vert \bar{m}_\perp \rangle
\label{eq:AX}
\eeq
and 
\beq
A \vert \psi \rangle = \vert \psi \rangle \qquad \mbox{for} \qquad 
\langle m_\pi \vert \psi \rangle = \langle m_\perp \vert \psi \rangle =0 .
\label{eq:AY}
\eeq

A unitary $A$ satisfying (\ref{eq:AW}),(\ref{eq:AX}) and 
(\ref{eq:AY}) is given by: 
\[
A = I + \Delta =
\]
\beq
I-  \vert m_\pi \rangle \langle m_\pi \vert -
\vert m_\perp \rangle \langle m_\perp \vert  + 
\vert \bar{m}_\pi \rangle \langle m_\pi \vert +
\vert \bar{m}_\perp \rangle \langle m_\perp \vert .
\label{eq:AZ}
\eeq
The perturbation $\Delta$ is the matrix
\beq
\Delta = (\vert \bar{m}_\pi \rangle- \vert m_\pi \rangle) 
\langle m_\pi \vert
+
(\vert \bar{m}_\perp \rangle - \vert m_\perp \rangle )
\langle m_\perp \vert .
\label{eq:BA}
\eeq
It is useful to express $\Delta$ directly in terms of 
the non-orthogonal states $\vert m_\pi \rangle$ 
and $\vert \bar{m}_\pi \rangle$: 
\[
\Delta =
\]
\beq
- \rho  (  \vert m_\pi \rangle - \vert \bar{m}_\pi \rangle )
(\langle m_\pi \vert - \langle \bar{m}_\pi \vert )
\label{eq:BB}
\eeq
where
\beq
\rho = {\cos (\theta) +1 \over \sin^2 (\theta) }.
\label{eq:BC}
\eeq
In this representation $\Delta$ is easily seen to be a 
rank-one Hermetian operator.  Using $\Delta+ \Delta^{\dagger} + 
\Delta \Delta^{\dagger}=0$
and 
$\langle m_{\pi} \vert (V_c-\bar{V}_c) \vert \bar{m}_{\pi} \rangle =  
\langle \bar{m}_{\pi} \vert (V_c-\bar{V}_c) \vert m_{\pi} \rangle = 0
$ gives the transformed confining interaction:
\[
V_c' = V_c+ \Delta M + M \Delta^{\dagger} +  \Delta M \Delta^{\dagger}
=
\]
\[
V+\rho 
( \vert m_\pi \rangle - \vert \bar{m}_\pi \rangle )
\langle \bar{m}_\pi \vert (V_c- \bar{V}_c) 
\]
\[
+ \rho (V_c- \bar{V}_c)\vert \bar{m}_{\pi} \rangle (\langle m_{\pi} \vert
- \langle \bar{m}_{\pi} \vert )  
\]
\beq
+ \rho^2 
( \vert m_\pi \rangle - \vert \bar{m}_\pi \rangle )
\langle \bar{m}_{\pi} \vert(V_c- \bar{V}_c)\vert \bar{m}_{\pi} \rangle
(\langle m_{\pi} \vert - \langle \bar{m}_{\pi} \vert ) .  
\label{eq:BD}
\eeq 
By construction the mass operator
$M' = M_0 + V_c'$:
\begin{itemize}
\item[a.)] Has the same spectrum as $M$.
\item[b.)] Has the same pion wave function as $\bar{M}$ 
\item[c.)] Differs from $M$ by the short range modification (\ref{eq:BD}) 
to the confining interaction.
\end{itemize}

Even though the transformation $A$ was constructed to transform 
the ground state wave function, it must also 
transform the $n>0$ states to preserve orthogonality.  The 
transformed states have the form 
\beq  
\vert {m}'_n \rangle = A \vert m_n \rangle =
\vert m_n \rangle + \rho
\langle \bar{m}_{\pi} \vert m_n \rangle 
\left (\vert m_{\pi} \rangle 
-  \vert \bar{m}_\pi \rangle \right ).
\label{eq:BDD}
\eeq

If the current $J^{\prime \mu}(x)$ is identified with the point
quark-antiquark impulse current, $J^{\mu}_p (x)$, then the
transformation $A$ generates exchange current contributions to the
current in the original representation:
\[
\langle p', m'_n \vert J_p^{\mu} \vert  p, m'_{l}
\rangle =
\]
\beq
\langle p', m_n \vert [ J_p^{\mu} + \Delta^{\dagger} J^{\mu}_{p} + 
J^{\mu}_{p} \Delta + 
\Delta^{\dagger} J^{\mu}_{p} \Delta] \vert  p, m_{l}
\rangle .
\label{eq:BE}
\eeq
Thus, form factors can be calculated using the current (\ref{eq:BE})
in the original representation or the point quark impulse 
current in the transformed representation (\ref{eq:BDD}).

This method provides an alternative means for describing 
electromagnetic properties of constituent quarks which replaces the  
constituent quark form factors by a different representation of 
the dynamics.

\section{Example} 

The simplicity of the construction is illustrated with an
example.  The constituent quark-antiquark model of Carlson,
Kogut and Pandharipande \cite{ckp} is taken as the constituent quark 
model that fits the meson spectrum.  The semi-relativistic Hamiltonian in
\cite{ckp} is interpreted as the mass operator in a fully 
relativistic model with the kinematic subgroup of the light front.
This relativistic model is described in the appendix.

The choice of kinematic subgroup does not effect the mass spectrum of 
the original model.  It does effect the value of the 
impulse current matrix elements.   

The mass operator in this model has the form
\beq
M = \sqrt{m_q^2 + k^2} + \sqrt{m_{\bar{q}}^2 + k^2} + V_c
\label{eq:CA}
\eeq
\beq
V_c (\lambda_1, \cdots , \lambda_4)  = 
\lambda_1 + {\lambda_2 \over r} + \lambda_3 r  + 
\delta \vec{s}_q \cdot \vec{s}_{\bar{q}} e^{-{r^2 \over 2\lambda_4^2} }  
\label{eq:CB}
\eeq
\beq
\delta = -  {\lambda_2 \over 3 m_q^2 \lambda_4^3 \sqrt{\pi}} .
\eeq
The model of \cite{ckp} also includes  spin-orbit and tensor interactions 
that vanish for the pion.  
The parameters $\lambda_1, \cdots, \lambda_4$ of the original Carlson
Kogut and Pandharipande model 
are listed on the first line of table 1.


\begin{center}
\begin{table} 
{\bf Table 1: Interaction Parameters } \\[1.0ex]
\begin{tabular}{|l|l|l|l|l|l|}
\hline
Interaction & $m\, (GeV)$ & $\lambda_1 \, (GeV)$ & $\lambda_2$ & $\lambda_3 \,(GeV)^2$ & $\lambda_4\, (GeV)^{-1}$   \\
\hline					      		      
$V_c $ &$.360$ &$-.777$  & $-.5$ &$.197$ & $.66$ \\
$\bar{V}_c$ &$.360$ &$-.911$  & $-.14$ &$.049$ & $.35$ \\
\hline
\end{tabular}
\label{tabspec}
\end{table}
\end{center}

The pion form factor can be expressed in terms of the
matrix elements of the $+$ component of the current in a frame where
the $+$ component of the momentum transfer vanishes\cite{fcwp2}
\cite{fcwp3}.  This matrix element is
evaluated with a point-like impulse current:
\beq
F_{\pi}(Q^2) =
\langle {q,\over 2} , m_\pi \vert \left ( J_{qp} ^{+}(0) + 
J_{\bar{q}p} ^{+} (0)\right  ) 
\vert -{q \over 2}, m_{\pi} \rangle
\label{eq:AR}
\eeq
where the states have the normalization
\beq
\langle p' , m_\pi  
\vert p, m_\pi \rangle = \delta (p^{+\prime} - p^+)
\delta^2 ( \vec{p}_{\perp}\,'- \vec{p}_{\perp}). 
\eeq
The calculation of the wave functions are done using 
the method described in \cite{kp}. If
the original Carlson, Kogut, Pandharipande wave pion function is used
in (\ref{eq:AR}), then the calculated form factor is larger than the
experimental form factor for both low \cite{amend} and high
\cite{volmer} momentum transfer.  The calculated form factor is given
by the dotted curves in figure 1 and figure 2.  

The parameters in the second line of table 1 define the interaction
$\bar{V}_c$.  The state $\vert \bar{m}_{\pi} \rangle$ is the ground
state of the mass operator $\bar{M}= M_0 + \bar{V}_c$.
The solid curves in figures 1 and 2 are the result of using $\vert
\bar{m}_\pi \rangle$ in (\ref{eq:AR}).  The calculated form factor
is consistent with the measured data, but the spectrum of 
the operator $\bar{M}$ no longer agrees with experiment. 

The method described in the previous section constructs a 
new mass operator, $M'$, that has the same spectrum as $M$ and 
the same pion wave function as $\bar{M}$. In this example the 
overlap parameter is
\beq
\cos (\theta) := \langle m_{\pi} \vert \bar{m}_{\pi} \rangle = .731 
\eeq
which gives $\rho = 3.71$.  The states $\vert
\bar{m}_{\pi} \rangle$ and $\vert m_\pi
\rangle$ and the overlap  $\langle \bar{m}_{\pi} \vert m_\pi
\rangle$ are used to define 
the unitary operator $A=I+\Delta$ using (\ref{eq:BB}).   
The transformed mass operator is $M' = M_0+
V_c'$ with $V_c'$ given by  (\ref{eq:BD}).  The operator $M'$ has the same 
mass spectrum as the original Carlson, Kogut, Pandharipande 
mass operator and has a pion eigenstate 
that can be used with the point impulse quark current to obtain 
the measured pion form factors.  By construction, the pion form 
factor in this model is given by the solid curves in
figures 1 and 2. 
Eigenstates of $M'$ for mesons 
other than the pion are obtained by applying $A$ to the corresponding 
eigenstates of $M$.

Figures 3 and 4 compare the coordinate (fig. 3) and momentum space
(fig. 4) wave functions of the pion for the mass operators $M$ (dotted
curve) and $M'$ (solid curve).  The Carlson, Kogut, and Pandharipande
wave functions have a smaller size in configuration space and more
high-momentum components than the wave-functions of the transformed
mass operator.  The form factor requires a more spread out wave
function in configuration space and a wave function with more low
momentum components in momentum space.  The transformation $A$ has the
overall effect of softening the wave functions of the original
Carlson, Kogut, and Pandharipande model.  This is consistent with the
calculations of Cardarelli et. al. \cite{salme}.  They use a similar
quark-antiquark model and need to introduce single quark form factors
to obtain measured pion form factors.  In \cite{salme} the quark form
factors also soften the wave function.

The pion decay constant, computed following \cite{fcwp3}, is $193$ MeV
in the light-front model with the original Carlson, Kogut,
Pandharipande parameters.  The transformation $A$ reduces this to 131
$MeV$, however it is still above the experimental value of 92 $MeV$.
This could be improved by considering more general transformations
$A$, or constituent quark models with smaller constituent quark masses
\cite{fcwp3}.

The overlap probability, $P=\vert \langle m_\pi \vert \bar{m}_\pi
\rangle \vert^2 = \cos^2 (\theta)$,  of the dotted and 
solid wave functions in fig. 3
and fig. 4 is .53.  This means that the overlap probability of
$\vert \bar{m}_{\pi} \rangle$ with the excited states of 
$M'$ is .47.  Figures 5 and 6 show the effect of the transformation 
$A$ on the point-like impulse form factors for the first radial 
excitation of the pion.  In these figures the dotted curve 
is the form factor in the original Carlson Kogut and Pandharipande 
model and the solid curve is the form factor in the transformed model.   
These plots indicate that the scattering equivalence also 
has significant effects on the excited mesons states with the 
same quantum numbers.   The form factors in the original and 
transformed models cross at about 1 $(GeV)^2$.  This is because 
the transformed excited state is not orthogonal to the 
original Carlson, Kogut, Pandharipande ground state, which has a 
significant amount of high momentum components.  The
operator $A$ constructed in this example is the simplest unitary 
operator that maps $\vert m_{\pi} \rangle$ to 
$\vert \bar{m}_{\pi} \rangle$.  Different choices of $A$ 
lead to different predictions for the form factors of 
the excited states. 
  
The effect of the transformation $A$ is to add an additional short
range structure to the original confining interaction.  Compared to
the original Carlson, Kogut, and Pandharipande interaction, the
additional short-ranged part contains non-localities.  In a
relativistic quantum theory there is no preferred reason to favor a
local over a non-local interaction except for mathematical simplicity.
Local and non-local interactions have similar strengths and defects.  In
relativistic models with short-ranged interactions, both the local and
non-local interactions are consistent with two-body cluster
properties, but both fail to be consistent with microscopic locality.
Microscopic derivations of local interactions necessarily make
implicit assumptions that lead to local interaction, however these
assumptions are not based on physics principles.  This shows that
there is no reason to consider the transformed interaction to be any
more or less fundamental than the original interaction.

\section{Conclusions} In this paper the freedom to change 
representation is used to construct a constituent quark model that fits
the meson mass spectrum and reproduces the pion form factor using only
point like constituent quarks.

This was done using scattering equivalences, which are unitary
operators of the form $A=I+ \Delta$, where $\Delta$ is asymptotically
zero \cite{wp}.  This paper utilized a limited class of these
operators where $\Delta$ is a finite rank operator on the internal
Hilbert space satisfying a domain restriction.  Desirable results were
achieved by considering only a small subset of rank-one $\Delta$'s.
Improvements to the fit of the pion form factor can be achieved by
considering a larger class of $\Delta$'s, such as compact $\Delta$'s.
Excellent results can be obtained with constituent quark models having
slighty lower constituent quark masses \cite{fcwp3}.  In this
application the quark masses cannot be used as parameters because they
appear in both the wave functions and the Clebsch-Gordon coefficients
of the Poincar\'e group.

The example in section three illustrates the simplicity and power of
the method. In this example the desired unitary scattering equivalence
is an easily constructed rank-one perturbation of the identity.
The only required input is the new state vector $\vert \bar{m}_{\pi}
\rangle$ and the overlap $\langle m_\pi \vert \bar{m}_{\pi} \rangle$.
The new representation does not require constituent quark/antiquark  
form factors.

The new model $M'$ is designed to be consistent with the meson spectrum
and the pion form factor.  For a given choice of $\Delta$ the model makes
predictions for elastic and transition form factors involving other
mesons which can (in principle) be tested against experiment. 
In addition, the transformed operators might be useful in 
formulating many-body constituent quark models of hadrons.   

The example started with relativistic constituent quark model with a
light-front kinematic symmetry, and produced a new relativistic model
with the same light-front kinematic symmetry, same mass spectrum,
where the pion form factor can be computed in point-quark impulse
approximation.  This example illustrates that constraints imposed by
the choice of kinematic subgroup and mass spectrum do not determine
the form factors.

Scattering equivalences are also known to exist \cite{wp} between
models with different kinematic subgroups.  These relationships can
also be exploited to construct equivalent current operators for models
with different forms of dynamics.  While the ambiguities in
representations of the dynamics are sometimes considered a liability,
this paper shows that they lead to a flexibility that can lead to a
simplification of the dynamics.


\section{Appendix}

\begin{appendix} 

The relativistic interpretation of the Carlson, Kogut, Pandharipande
semi-relativistic Hamiltonian is discussed in this appendix.

A relativistic quantum mechanical model is defined by 
a unitary representation of the Poincar\'e group acting 
on a model Hilbert space. The mass operator of this 
representation is necessarily a dynamical operator.

A representation of the quark-antiquark Hilbert space and a
representation of the Poincar\'e Lie algebra where the mass operator
is the semi-relativistic Carlson, Kogut, and Pandharipande Hamiltonian
is exhibited in this section.  The generators are chosen so the
generators of transformations that leave the light front $x^+=0$
invariant are kinematic.  This defines a relativistic light-front 
dynamics.

The Hilbert space in this model is the tensor product of the mass $m_q$ spin ${1
\over 2}$ irreducible representation of the Poincar\'e group
associated with the quark and the mass $m_{\bar{q}}$ spin ${1 \over
2}$ irreducible representation of the Poincar\'e group associated with
the antiquark.  The light-front components of the 
single particle momenta $p_i:= (p^+_i, p^1_i, p^2_i)
= (p^+_i, \vec{p}_{i \perp} )$ and the 3-component of the 
light front spin $j^3_i$ are a maximal set of commuting observables on the 
single quarks spaces.  The model Hilbert space is the tensor product of 
the space of square integrable functions of $(p_i, \mu_i)$;
\beq 
\psi_q (p_q, \mu_q) \otimes \psi_{\bar{q}} (p_{\bar{q}} , \mu_{\bar{q}})
\label{eq:DA}
\eeq

The tensor product of irreducible representations of the 
Poincar\'e group is reducible.  The Clebsch-Gordon 
coefficients of the Poincar\'e group are \cite{bkwp}
\[
\langle p_q, \mu_q, p_{\bar{q}}, \mu_{\bar{q}} 
\vert P, \mu (k,j; l,s ) \rangle
=
\]
\[
\delta (P-p_1 -p_2) {\delta (k -k(p_1,p_2)) \over \vert k^2 \vert} 
\sqrt{{ \omega_1 (k) \omega_2 (k) P^+ \over p_1^+ p_2^+ 
m(k)}} 
\]
\[ 
\times D^{1/2}_{\mu_1 \mu_1'} [B_f^{-1} (k_1) B_c (k_1)] 
D^{1/2}_{\mu_2 \mu_2'} [B_f^{-1} (k_2) B_c (k_2)] Y^l_{\mu_l} (\hat{k} )
\]
\beq
\times \langle {1 \over 2}, \mu_1,{1 \over 2} , \mu_2 \vert s, \mu_s \rangle
\langle l, \mu_l, s, \mu_s \vert j, \mu \rangle  
\label{eq:DB}
\eeq
where 
\beq
m=m(k)= \sqrt{k^2 +m_q^2} + \sqrt{k^2 +m_{\bar{q}}^2}
\label{eq:DC}
\eeq
is replaced by the kinematic relative momentum, $k$.
$B_f (p)$ and $B_c (p)$ are light-front and canonical 
Lorentz boosts respectively, and the momenta $k_i$ are defined by
\beq
k_i = B^{-1}_f (P) p_i 
\label{eq:DD}
\eeq
The discrete indices $l,s$ label multiple copies of the same irreducible
representation that appear in the tensor product.  From the structure 
of the Clebsch-Gordon coefficient above it is obvious that $s\in {0,1}$ and
$\vert j-s \vert \leq l \leq j+s$.  

The Clebsch-Gordon coefficients lead to a representation of the quark
antiquark Hilbert space as the space of square integrable functions of
the variables $( P^+,\vec{P}_{\perp} , \mu , k,j, l,s )$.  Vectors 
in the Hilbert space are square integrable functions of theses variables
\beq
\psi (P,\mu,k,j,l,s) = \langle   P^+,\vec{P}_{\perp} , \mu , k,j, l,s
\vert \psi \rangle.
\label{eq:DE}
\eeq

The Carlson, Kogut and Pandharipande semi-relativistic Hamiltonian is 
embedded in this space as follows:
\[
\langle P, \mu (k,j;l,s) \vert M \vert P', \mu' (k',j';l',s')
\rangle =
\]
\beq 
\delta (P-P') \delta_{jj'} \delta_{\mu \mu'} 
\langle k,l,s \Vert M^j \Vert k',l',s' \rangle 
\label{eq:DF}
\eeq
This defines $M$ in terms of the kernel, 
$\langle k,l,s \Vert M^j \Vert k',l',s' \rangle $,
of the Carlson, Kogut and Pandharipande semi-relativistic Hamiltonian.   
This kernel is identified with the semi-relativistic Hamiltonian 
given by (\ref{eq:CA}) and (\ref{eq:CB}).

Given this mass operator the following operators \cite{bkwp}
\beq
\vec{E}= -i P^+ {\partial \over \partial \vec{P}_{\perp}}  
\qquad
K_3= -i P^+ {\partial \over \partial P^+}  
\label{eq:DH}
\eeq
\beq
J_3= j_3 - {1 \over P^+} \hat{z} \cdot ( \vec{P} \times \vec{E} )
\label{eq:DI}
\eeq
\beq
P_- = {\vec{P}_{\perp} \cdot \vec{P}_{\perp}  - M^2 \over P^+}
\label{eq:DJ}
\eeq
\[
\vec{J}_{\perp} =
\]
\beq
{1 \over P^+}\left ( {P^+ - P^-  \over 2} (\hat{z} \times \vec{E} )  
- (\hat{z} \times \vec{P} )K_3 + \vec{P}_{\perp}j_3 + 
M\vec{j}_{\perp}\right )
\label{eq:DK}
\eeq
along with  the multiplication operators 
$P^+$ and $\vec{P}_{\perp}$, define a set of self-adjoint operators 
on the space of square integrable functions of
$\{ P,\mu,k,j,l,s\}$ that satisfy the Poincar\'e Lie algebra
with the Carlson Kogut and Pandharipande mass operator (\ref{eq:DF}),
(\ref{eq:CA}) and (\ref{eq:CB}).

The interaction in the above expressions appears in both $P^-$ and
$\vec{J}_{\perp}$.  The remaining generators, which generate
transformations that leave the light front invariant, do not involve
$M$ thus are kinematic.  It is clear that the same properties hold if $M$
is replaced by $M'$.  Unlike Fock space motived models
\cite{bertrand}, 
the Poincar\'e symmetry is exact.

Note that to compute current matrix elements with an impulse current
it is necessary to use the Clebsch Gordon coefficients (\ref{eq:DB})
to transform the eigenstates of $M$ and $M'$ to single quark
variables.  Because the Clebsch Gordon coefficients depend on the
quark masses, (\ref{eq:DC}),  the quark masses were not allowed to
vary in determining $\bar{M}$.  

\end{appendix}

 
 
 

\vfill\eject 

\begin{figure*}
\rotatebox{00}{\resizebox{4.0in}{!}{
\includegraphics{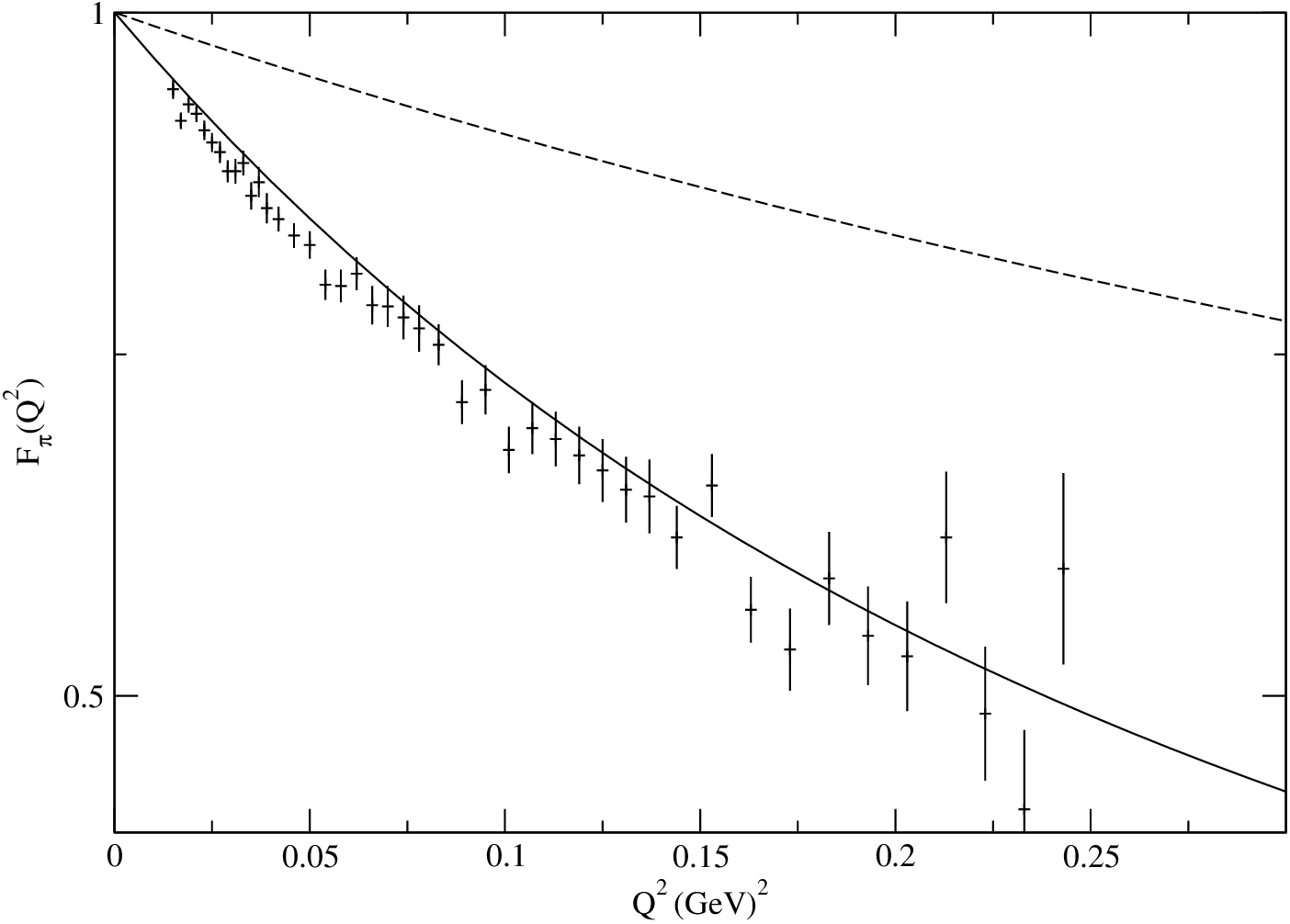}}}
\caption{\label{fig:a}Pion form factor at low $Q^2$.}
\end{figure*}

\vfill\eject 

\begin{figure*}
\rotatebox{00}{\resizebox{4.0in}{!}{
\includegraphics{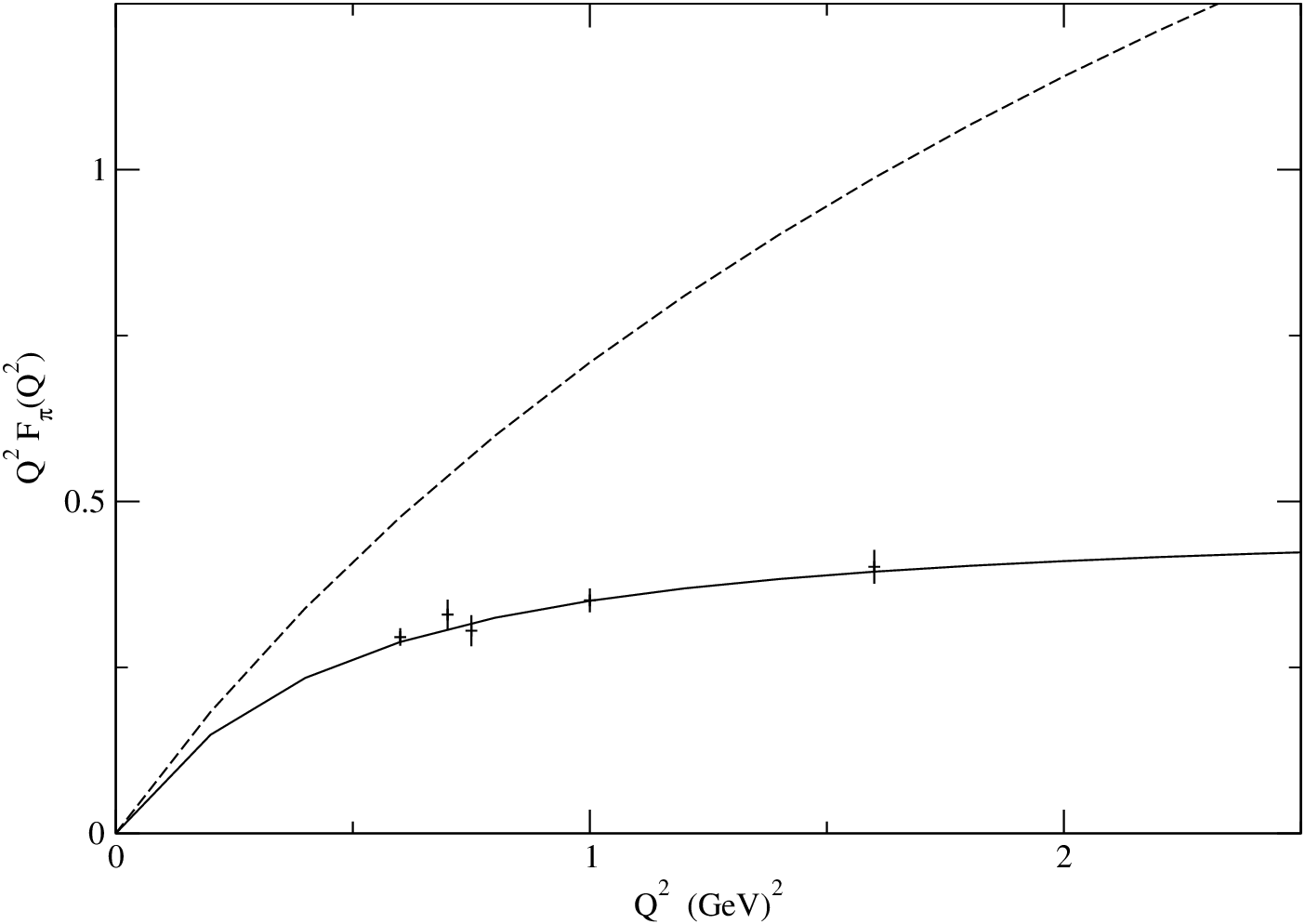}}}
\caption{\label{fig:b}Pion form factor at high $Q^2$.}
\end{figure*}

\vfill\eject 

\begin{figure*}
\rotatebox{00}{\resizebox{4.0in}{!}{
\includegraphics{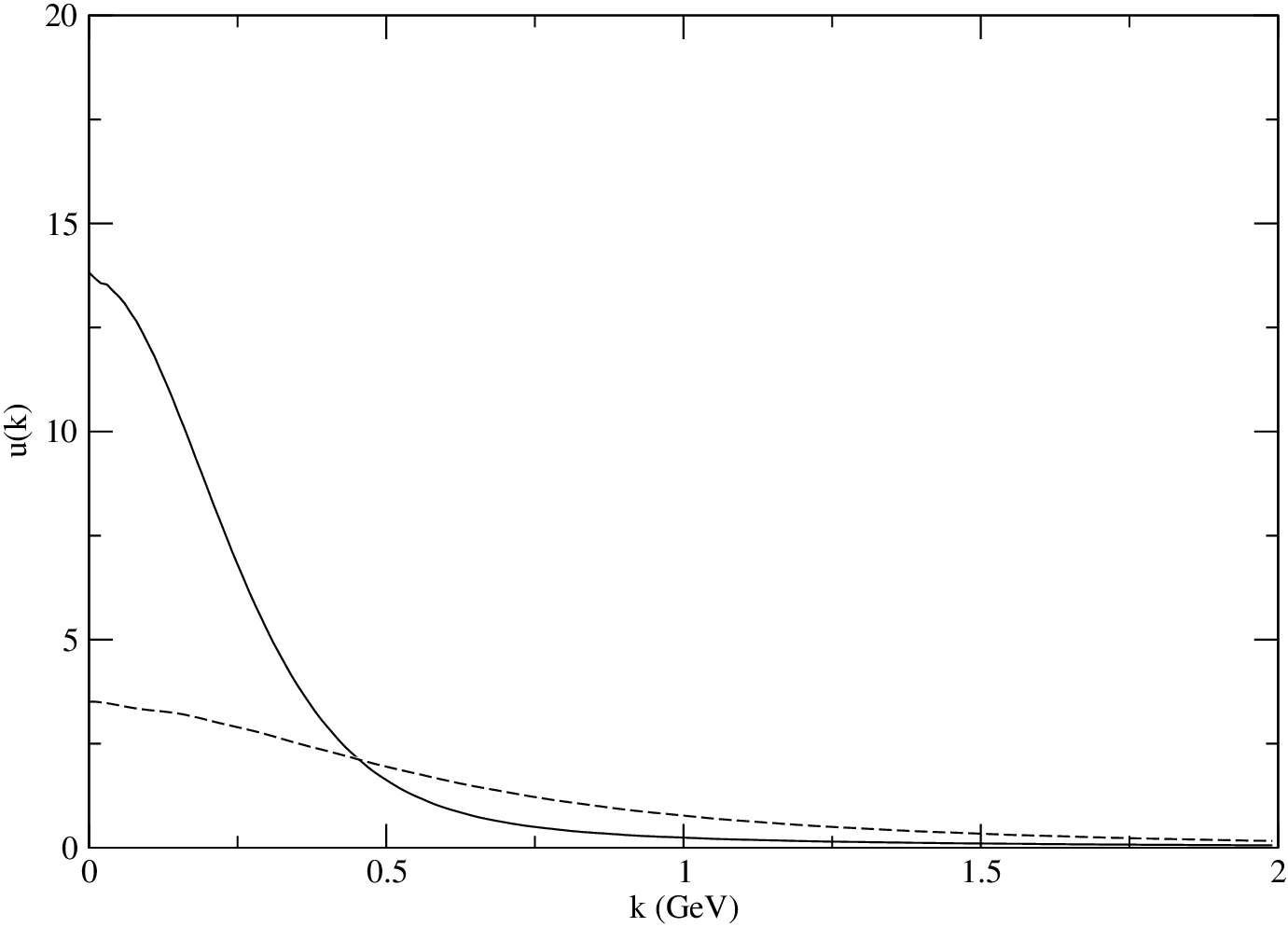}}}
\caption{\label{fig:c}Comparison of k-space wave functions.}
\end{figure*}

\vfill\eject 

\begin{figure*}
\rotatebox{00}{\resizebox{4.0in}{!}{
\includegraphics{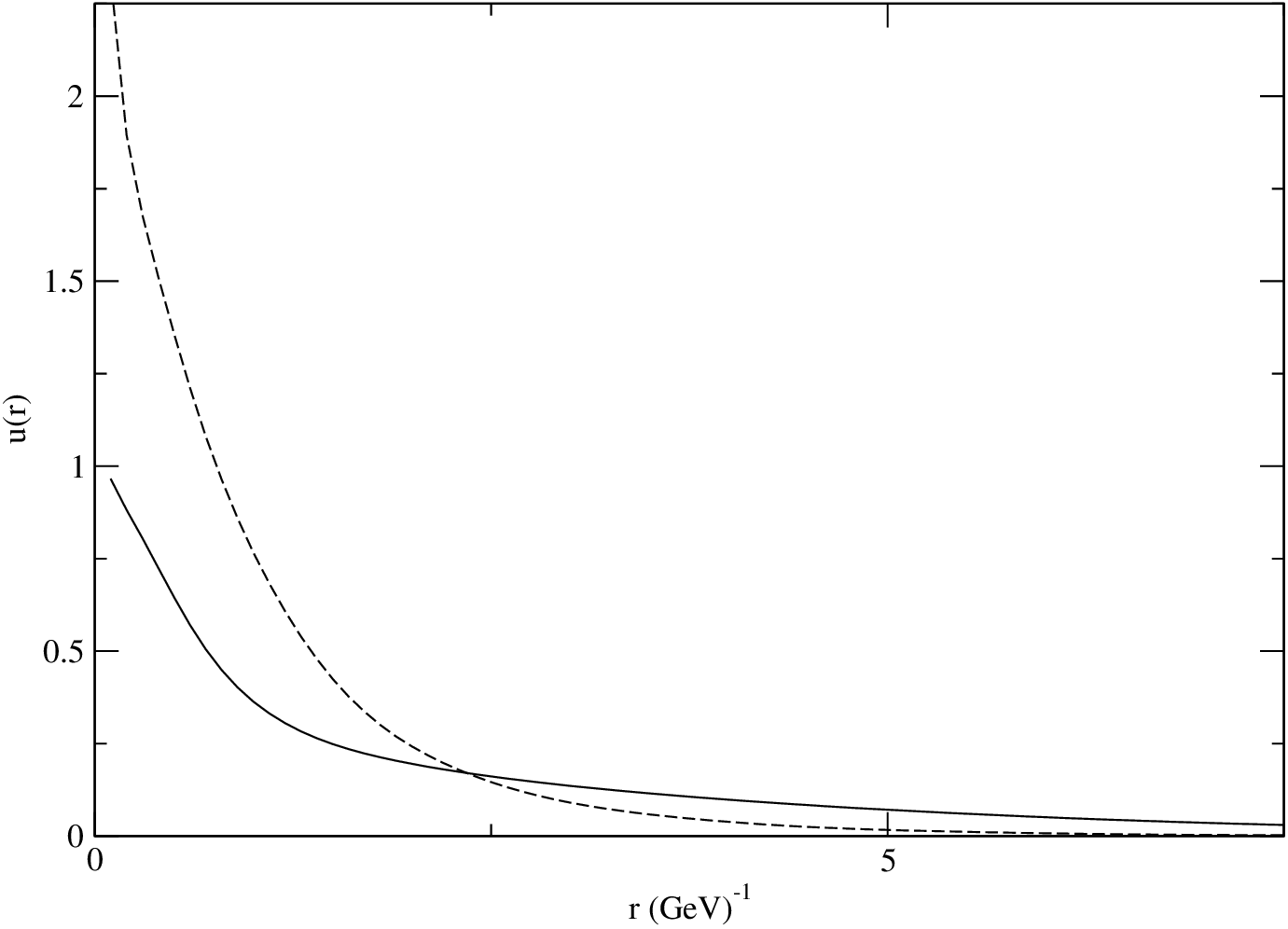}}}
\caption{\label{fig:d}Comparison of r-space wave functions.}
\end{figure*}

\begin{figure*}
\rotatebox{00}{\resizebox{4.0in}{!}{
\includegraphics{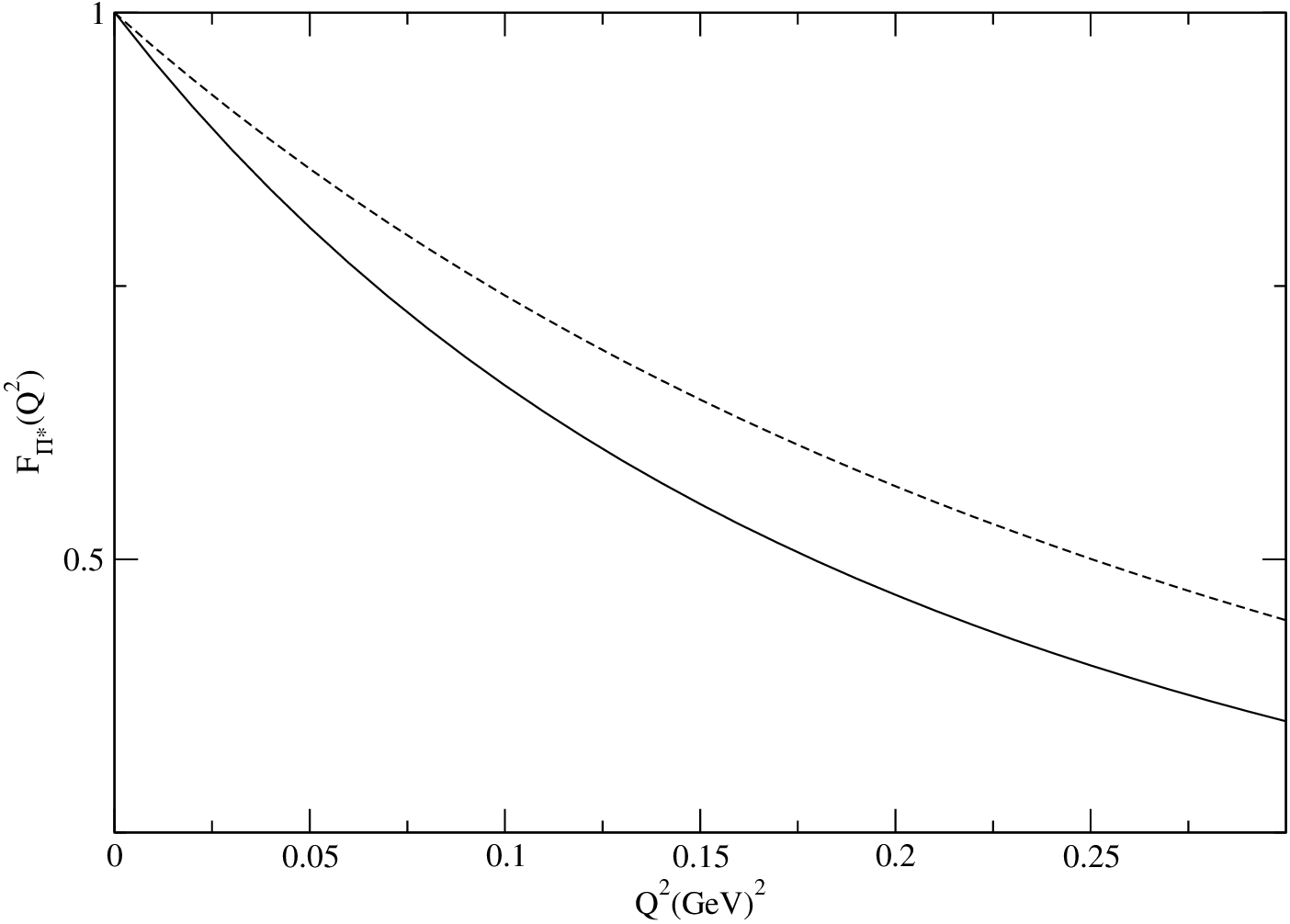}}}
\caption{\label{fig:e}$\pi^*$ form factor at low $Q^2$.}
\end{figure*}

\begin{figure*}
\rotatebox{00}{\resizebox{4.0in}{!}{
\includegraphics{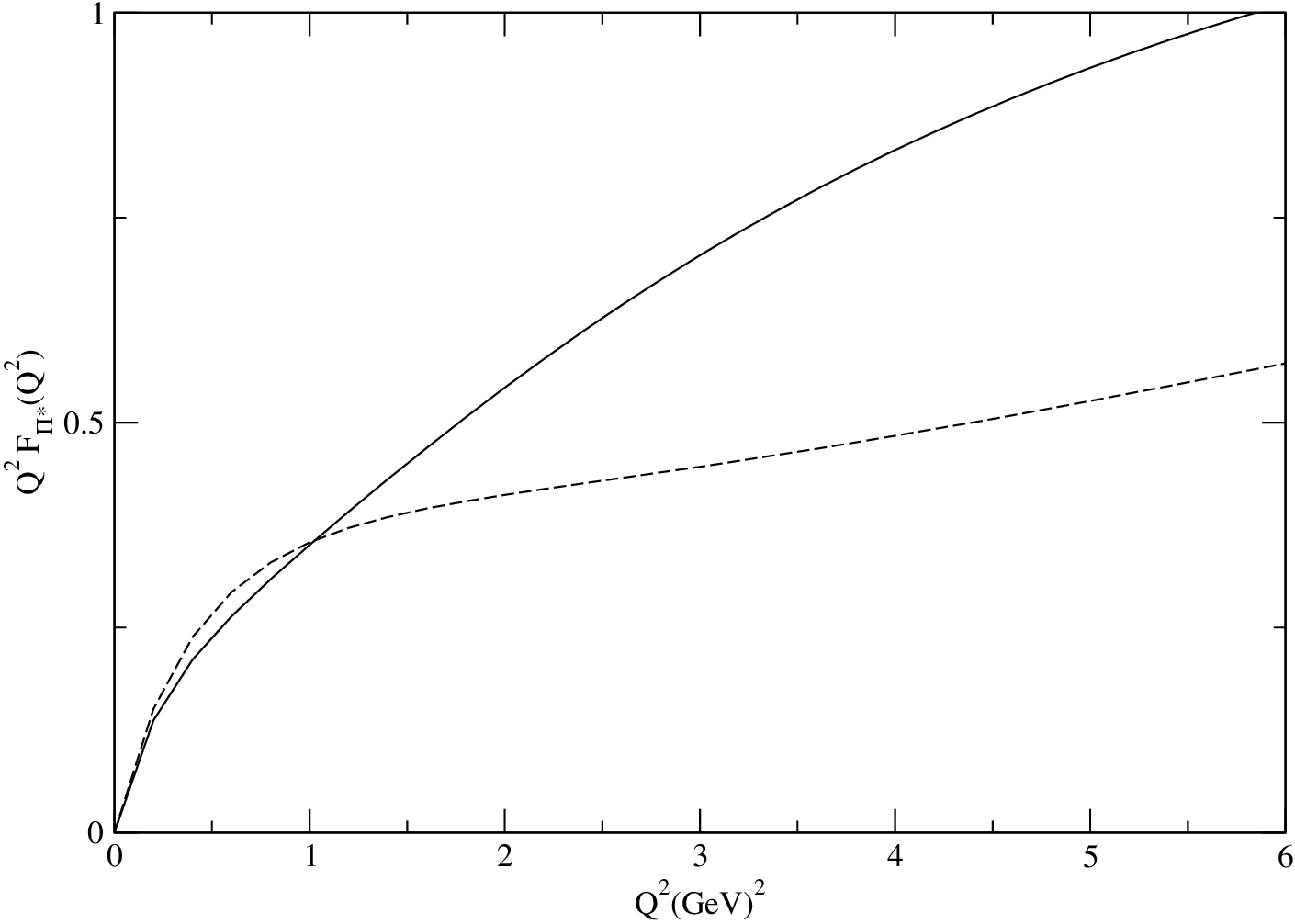}}}
\caption{\label{fig:f}$\pi^*$ form factor at high $Q^2$.}
\end{figure*}


\begin{thebibliography}{99}
\bibitem{ckp} J. Carlson, J. B. Kogut, V.R. Pandharipande, Phys. Rev. 
D 28, 2807 (1983) 
\bibitem{bkwp} B.D. Keister and W.N. Polyzou, Adv. Nuc. Phys. {\bf 20}, 225(1991)
\bibitem{fc} F. Coester, Prog. Part. Nucl. Phys.{\bf 29} 1 (1992)
\bibitem{salme}F. Cardarelli et al. Phys. Lett. B {\bf 332} 1 (1994); 
{\bf 357} 267 (1995)
\bibitem{allen}T.W. Allen and W.H. Klink,  Phys. Rev. C {\bf 58} , 3670 (1998)
\bibitem{fcdr} B. Julia-Dia, F. Coester and D. O. Riska, Phys. 
Rev. C69:035212,2004 
\bibitem{bertrand1}B. Desplanques, nucl-th/0405060
\bibitem{krutov} A. F. Krutov and V. E. Troitsky, 
Theor. Math. Phys.  116, 219 (1998),  hep-ph/9805312
\bibitem{sw} R. F. Streater and A. S. Wightman, PCT and Spin and Statistics, and all that, W. A. Benjamin, 1964. 
\bibitem{dick} R. J. Furnstahl, H. W. Hammer, Negussie Tirfessa,
Nucl. Phys. A689:846-868,2001 
\bibitem{fcwp1} F. Coester and W. N. Polyzou, Phys. Rev. D26, 1348(1982). 
\bibitem{wp} W. N. Polyzou, J. Math. Phys. 43, 6024 (2002)
\bibitem{fcwp2} P. L. Chung, F. Coester and W. N. Polyzou, 
Phys. Lett. B {\bf 205}, 545 (1988)
\bibitem{fcwp3} F. Coester and W. N. Polyzou, Lanl AxXiv nucl-th/0405082
\bibitem{kp}B. D. Keister and W. N. Polyzou, J. Comp. Phys. {\bf 134} 
231 (1997)
\bibitem{amend} S. R. Amendolia et al. Nuc. Phys. B  {\bf 277}, 168 (1986)
\bibitem{volmer} J. Volmer et al. Phys. Rev. Lett. {\bf 86} , 1713  (2001)
\bibitem{bertrand} A. Amghar, B. Desplanques and L.Theussl, 
Phys. Lett. B {\bf 574} , 201 (2003)
\end{thebibliography}
\end{document}